\documentclass[final,5p,times,twocolumn,numbers]{elsarticle}

\usepackage{amssymb}
\usepackage{hyperref}
\makeatletter
\def\ps@pprintTitle{%
  \let\@oddhead\@empty
  \let\@evenhead\@empty
  \let\@oddfoot\@empty
  \let\@evenfoot\@oddfoot
}
\makeatother

\usepackage{lipsum}

\usepackage{transparent}
\usepackage{algorithm}
\usepackage{algpseudocode}

\usepackage{hyperref}
\hypersetup{
    colorlinks=true,
    linkcolor=blue,
    filecolor=magenta,      
    urlcolor=cyan,
    pdftitle={Overleaf Example},
    pdfpagemode=FullScreen,
    }

\usepackage{comment}
\usepackage{listings}

\usepackage[tight,footnotesize]{subfigure}
\usepackage{color}
\usepackage{tikz}
\usepackage{pgfplots}

\usepackage{xcolor}

\definecolor{codegreen}{rgb}{0,0.6,0}
\definecolor{codegray}{rgb}{0.5,0.5,0.5}
\definecolor{codepurple}{rgb}{0.58,0,0.82}
\definecolor{backcolour}{rgb}{0.95,0.95,0.92}

\lstdefinestyle{mystyle}{
    backgroundcolor=\color{backcolour},   
    commentstyle=\color{codegreen},
    keywordstyle=\color{magenta},
    numberstyle=\tiny\color{codegray},
    stringstyle=\color{codepurple},
    basicstyle=\ttfamily\footnotesize,
    breakatwhitespace=false,         
    breaklines=true,                 
    captionpos=b,                    
    keepspaces=true,                 
    numbers=left,                    
    numbersep=5pt,                  
    showspaces=false,                
    showstringspaces=false,
    showtabs=false,                  
    tabsize=2
}
\usepackage{stfloats}

\usepackage{capt-of}

\begin{document}

\renewcommand{\labelenumii}{\arabic{enumi}.\arabic{enumii}}

\begin{frontmatter}

\title{TensorConvolutionPlus: A python package for distribution system flexibility area estimation}

\author[first]{Demetris~Chrysostomou, José~Luis~Rueda~Torres, and~Jochen~Lorenz~Cremer}
\affiliation[first]{organization={TU Delft}, Department of Electrical Sustainable Energy,
            city={Delft},
            country={The Netherlands}}

\begin{abstract}
Power system operators need new, efficient operational tools to use the flexibility of distributed resources and deal with the challenges of highly uncertain and variable power systems. Transmission system operators can consider the available flexibility in distribution systems (DSs) without breaching the DS constraints through flexibility areas. However, there is an absence of open-source packages for flexibility area estimation. This paper introduces TensorConvolutionPlus, a user-friendly Python-based package for flexibility area estimation. The main features of TensorConvolutionPlus include estimating flexibility areas using the TensorConvolution+ algorithm, the power flow-based algorithm, an exhaustive PF-based algorithm, and an optimal power flow-based algorithm. Additional features include adapting flexibility area estimations from different operating conditions and including flexibility service providers offering discrete setpoints of flexibility. The TensorConvolutionPlus package facilitates a broader adaptation of flexibility estimation algorithms by system operators and power system researchers.
\end{abstract}

\begin{keyword}
Python package \sep flexibility area estimation \sep distribution system flexibility \sep TSO-DSO coordination 

\end{keyword}

\end{frontmatter}

\section*{Metadata}

\noindent\begin{minipage}{\textwidth}
\centering
\begin{tabular}{|l|p{7.5cm}|p{9.1cm}|}
\hline
\textbf{Nr.} & \textbf{Code metadata description} & \textbf{Metadata} \\
\hline
C1 & Current code version & v0.1.1 \\
\hline
C2 & Permanent link to code/repository used for this code version & \url{https://github.com/Demetris-Ch/TensorConvolutionFlexibility} \\
\hline
C3 & Legal Code License   &  CC-BY \\
\hline
C4 & Code versioning system used & git \\
\hline
C5 & Software code languages, tools, and services used & Python \\
\hline
C6 & Compilation requirements, operating environments \& dependencies & Python$\geq3.10$, matplotlib$\geq3.8.2$, networkx$\geq3.1$, numpy$\geq1.24.3$, pandapower$\geq2.13.1$, pandas$\geq1.5.3$, scikit-learn$\geq1.3.0$, scipy$\geq1.11.2$, seaborn$\geq0.13.2$, tntorch$\geq1.1.1$, torch$\geq2.0.1$, tqdm$\geq4.66.1$ \\
\hline
C7 & Link to developer documentation/manual & \url{https://demetris-ch.github.io/TensorConvolutionFlexibility/} \\
\hline
C8 & Support email for questions & D.Chrysostomou@tudelft.nl\\
\hline
\end{tabular}
\captionof{table}{Code metadata}
\label{codeMetadata} 
\end{minipage}

\clearpage

\section{Motivation and significance}
\label{mas}
Power systems encounter an operational transition as renewable energy sources (RES) penetration rises, and the conventional generation output decreases. This operational transition includes coordinating transmission system operators (TSOs) and distribution system operators (DSOs). RES are mainly connected to distribution systems (DSs) and have high variability and uncertainty, challenging the TSOs and DSOs who need to maintain their system balance. However, RES and active users in DSs can also offer flexibility to contribute to the reduction of these challenges. This flexibility corresponds to the RES or active users changing their generation or consumption setpoints to support the system operators. The RES and active users that offer flexibility constitute the flexibility service providers (FSPs). 

Flexibility areas (FAs) are areas in the active (P) and reactive (Q) power plane, illustrating which setpoints TSOs can achieve at a TSO-DSO interconnection node when utilizing feasible flexibility from the DSs.

FA estimation approaches mainly apply power flows (PF) or optimal power flows (OPF)  \cite{silva2018challenges, capitanescu2018tso, kalantar2019characterizing, savvopoulos2021contribution, gonzalez2018determination, 10083112, 10202983, 10202791, chen2021enhancing, bolfek2021analysis, chrysostomou2023exploring} to explore the limits of the offered flexibility in the PQ space. PF-based algorithms are simple and consistent but slow, whereas OPF-based algorithms are faster but can have convergence issues. A recently proposed FA estimation algorithm, TensorConvolution+ \cite{10663439}, efficiently explores the limits and the density of feasible flexibility shift combinations to reach each FA setpoint. TensorConvolution+ applies convolution and tensor operations to combine flexibility shifts and evaluate their feasibility for the system's technical constraints. Additional functionalities of the TensorConvolution+ approach include storing tensors from prior estimations and adapting FAs for altered operating conditions (OCs). Estimating the density of feasible combinations (DFC) for each FA setpoint through alternative algorithms is only functional if the PF algorithms perform PF for each possible flexibility shift combination, i.e., exhaustive search.

The operational transition and data availability in power systems provided opportunities for digitalizing power systems. This digitalization corresponds to more intelligent, effective, green power grid operations \cite{cali2021introduction}. The main drivers for change in power systems are decarbonization, digitalization, and decentralization, with flexibility as a key for decarbonization \cite{DISILVESTRE2018483}. The digitalization of power systems resulted in the emergence of open-source tools. Power systems open-source tools include the PandaPower \cite{thurner2018pandapower} in Python, PSAT \cite{milano2005open} in Matlab and GNU/Octave, MatPower \cite{5491276} in Matlab. More recent tools with increased efficiency include \cite{10267233} in C++. As highlighted by \cite{thurner2018pandapower}, software developed in languages with open-source licenses, such as Python, C++, and  Julia, can be used as stand-alone or extended with other libraries. These advantages of open-source libraries drove researchers to design more specialized power system-related packages such as \cite{JOHNSTON2019100251, MIRZ2019100253, PLIETZSCH2022100861, ramos2022opentepes}. FA estimation is an emerging field in power engineering that can improve the power system stability and utilization of flexibility from decentralized resources. However, currently, there are no open-source FA estimation packages. An open-source package for FA estimation can accelerate the adoption of FAs by power system operators and attract more researchers to this emerging field.

The developed Python-based package for FA estimation focuses on the TensorConvolution+ algorithm \cite{10663439} but also includes a traditional PF-based algorithm, an exhaustive PF-based algorithm, and an OPF-based algorithm. 

\section{Software description} \label{sec:sd}

The software framework is implemented in Python. The package can be installed from the Python package index (PyPi). The code implementation is available on GitHub. The documentation for the package's main functions, classes, supporting functions, and case studies is available online and was built using the Sphinx library \cite{brandl2010sphinx}.  

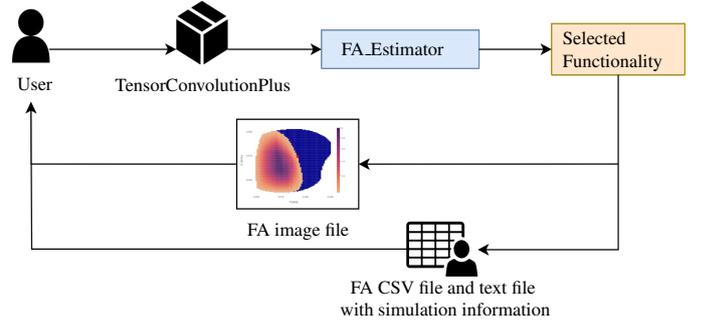
\begin{figure}[!tb]
\centerline{\def\svgwidth{260bp}
\fontsize{7pt}{7pt}\selectfont
\begingroup%
  \makeatletter%
  \providecommand\color[2][]{%
    \errmessage{(Inkscape) Color is used for the text in Inkscape, but the package 'color.sty' is not loaded}%
    \renewcommand\color[2][]{}%
  }%
  \providecommand\transparent[1]{%
    \errmessage{(Inkscape) Transparency is used (non-zero) for the text in Inkscape, but the package 'transparent.sty' is not loaded}%
    \renewcommand\transparent[1]{}%
  }%
  \providecommand\rotatebox[2]{#2}%
  \newcommand*\fsize{\dimexpr\f@size pt\relax}%
  \newcommand*\lineheight[1]{\fontsize{\fsize}{#1\fsize}\selectfont}%
  \ifx\svgwidth\undefined%
    \setlength{\unitlength}{422.55496216bp}%
    \ifx\svgscale\undefined%
      \relax%
    \else%
      \setlength{\unitlength}{\unitlength * \real{\svgscale}}%
    \fi%
  \else%
    \setlength{\unitlength}{\svgwidth}%
  \fi%
  \global\let\svgwidth\undefined%
  \global\let\svgscale\undefined%
  \makeatother%
  \begin{picture}(1,0.44289965)%
    \lineheight{1}%
    \setlength\tabcolsep{0pt}%
    \put(0,0){\includegraphics[width=\unitlength,page=1]{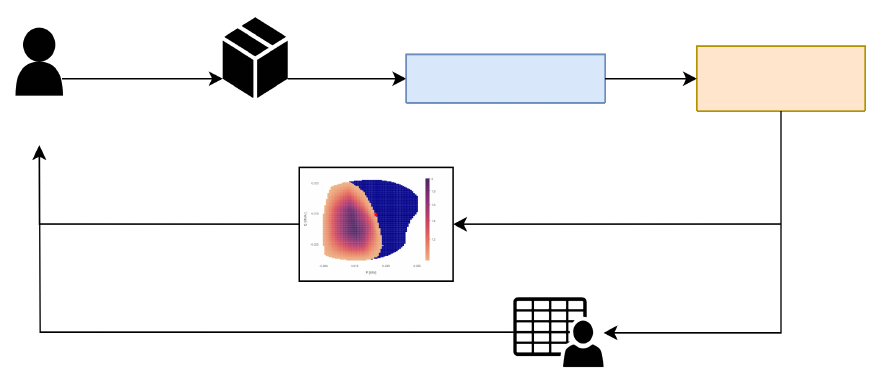}}%
    \put(0.48959142,0.34584522){\color[rgb]{0,0,0}\makebox(0,0)[lt]{\lineheight{1.25}\smash{\begin{tabular}[t]{l}FA\_Estimator\end{tabular}}}}%
    \put(0.8068345,0.36139634){\color[rgb]{0,0,0}\makebox(0,0)[lt]{\lineheight{1.25}\smash{\begin{tabular}[t]{l}Selected\\Functionality\end{tabular}}}}%
    \put(0.02409423,0.29504484){\color[rgb]{0,0,0}\makebox(0,0)[lt]{\lineheight{1.25}\smash{\begin{tabular}[t]{l}User\end{tabular}}}}%
    \put(0.16478924,0.2940081){\color[rgb]{0,0,0}\makebox(0,0)[lt]{\lineheight{1.25}\smash{\begin{tabular}[t]{l}TensorConvolutionPlus\end{tabular}}}}%
    \put(0.35451304,0.08562296){\color[rgb]{0,0,0}\makebox(0,0)[lt]{\lineheight{1.25}\smash{\begin{tabular}[t]{l}FA image file\end{tabular}}}}%
    \put(0.50271447,0.00061013){\color[rgb]{0,0,0}\makebox(0,0)[lt]{\lineheight{1.25}\smash{\begin{tabular}[t]{l}FA CSV file and text file\end{tabular}}}}%
    \put(0.48771447,-0.03061013){\color[rgb]{0,0,0}\makebox(0,0)[lt]{\lineheight{1.25}\smash{\begin{tabular}[t]{l}with simulation information\end{tabular}}}}%
  \end{picture}%
\endgroup%
}
    \caption{TensorConvolutionPlus package usage through the script  (\protect\includegraphics[height=1.0em]{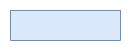}) FA\_Estimator and its main functionalities (\protect\includegraphics[height=1.0em]{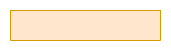}).}
    \label{fig:usage}
\end{figure}

The seven main software functionalities are two PF-based algorithms, one OPF-based algorithm, and four versions of the TensorConvolution+ algorithm. 
Fig.\ref{fig:usage} illustrates the usage of the algorithm. 
The user calls the \textit{FA\_Estimator} script of the TensorConvolutionPlus package and selects one of the main functionalities. The selected algorithm functionality estimates the FA and stores locally:
\begin{enumerate}
    \item The FA image in a portable document format (PDF) file.
    \item The FA results in a comma-separated values (CSV) file.
    \item A text file with the simulation information on duration and algorithm-specific details.
\end{enumerate}
The \textit{tcp\_plus\_save\_tensors} also includes additional files from the FA results. The user inputs depend on the functionality.

\subsection{Software architecture}
The proposed software architecture intends to allow efficient modification and expansion of specific sub-processes of the FA estimation problem. Tab.\ref{tab:1} highlights the roles of the Python scripts implementing the package functionalities. Users can identify which script and functionalities to modify or expand to fulfill additional needs. For example, to add or modify the plotting functions of the package, one would modify the \textit{plotting} script. For more complex modifications, such as adding different sampling techniques for the PF-based algorithms, the \textit{data\_sampler} and \textit{json\_reader} would be the only scripts requiring modification. The \textit{json\_reader} modification corresponds to adding new acceptable options (e.g., new sampling distribution). An input outside the acceptable options would stop the process before the simulation to avoid erroneous or untested results. This architecture also allows potential future expansions to new FA estimation algorithms, where an additional script can be created and integrated with the \textit{json\_reader} and \textit{FA\_Estimator} scripts without impacting other processes.

\noindent
\begin{table*}[!t]
\centering
  \begin{tabular}{ll}
    \hline
    Script & Role \\
    \hline
    FA\_Estimator & Package main script which includes the main functionalities.\\
    json\_reader & (i) Read input settings and create a SettingReader object with the algorithm parameters.\\
        & (ii) Validate that the inputs are within the acceptable options.\\
    data\_sampler & Sample flexibility shifts from flexibility providers.\\
    scenario\_setup & Update network and SettingReader object based on the algorithm input parameters.\\
    opf & Perform the OPF-based FA estimation algorithm.\\
    monte\_carlo & Perform the PF-based FA estimation algorithms. \\
    conv\_simulations & Perform the TensorConvolution+ algorithm functionalities.\\
    utils &  Provide generic functions to the other scripts.\\
    plotting & Generate figures of resulting FAs.\\
    \hline
  \end{tabular}
  \label{tab:1}
  \caption{Package script roles.}
\end{table*}
\noindent

\begin{figure}[!tb]
\centerline{\def\svgwidth{260bp}
\fontsize{7pt}{7pt}\selectfont
    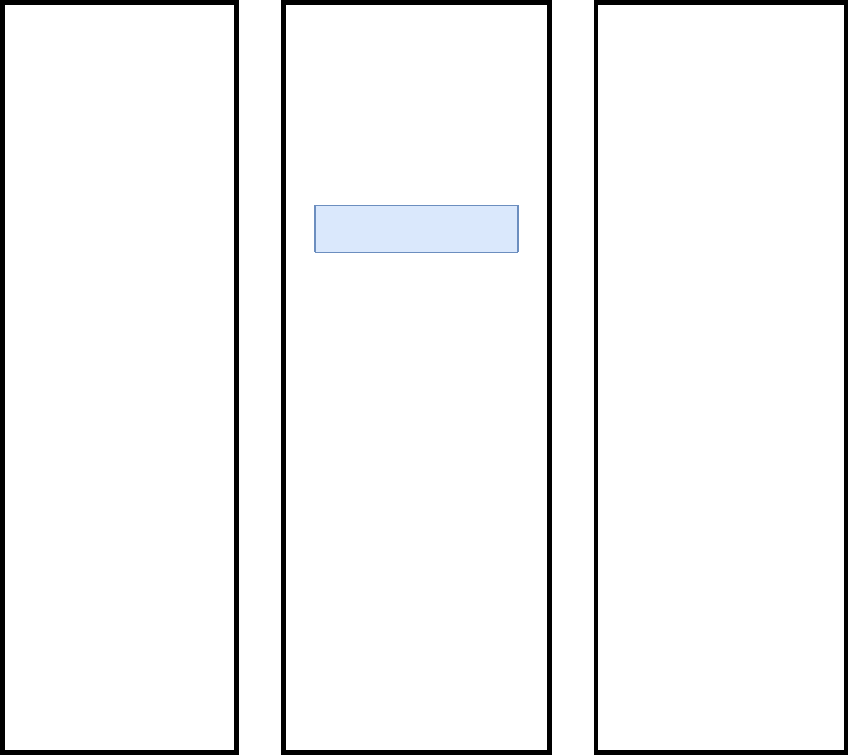}
    \caption{Package script (\protect\includegraphics[height=1.0em]{images/script.png}) dependencies to python standard library (\protect\includegraphics[height=1.0em]{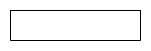}) and external libraries (\protect\includegraphics[height=1.0em]{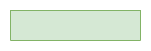}). The (\protect\includegraphics[height=1.2em]{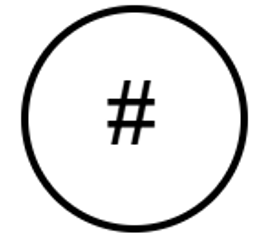}) indicates the number of scripts using each library.}
    \label{fig:libs}
\end{figure}

The package's GitHub repository includes the Python scripts under the "src/TensorConvolutionPlus" directory. Fig.\ref{fig:libs} shows the package scripts and their dependencies on the Python standard library and other external libraries. The Python standard library dependencies do not require installation besides Python. The external libraries require installation. The user can install all dependencies by installing the \textit{TensorConvolutionPlus} package.  

 \subsection{Software functionalities} \label{sec:sfn}
\begin{figure}[!tb]
\centerline{\def\svgwidth{275bp}
\fontsize{7pt}{7pt}\selectfont
    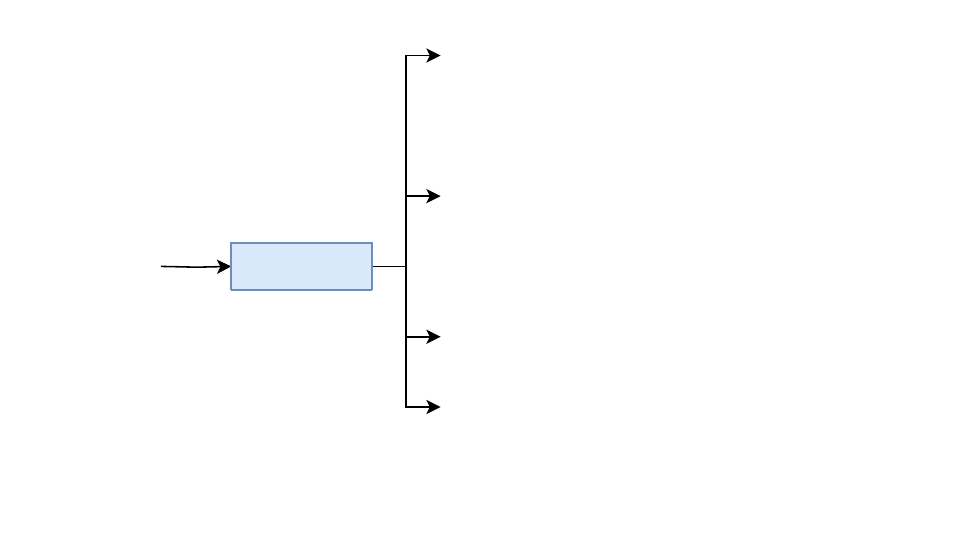 }
    \caption{Package main functions (\protect\includegraphics[height=1.0em]{images/func.png}) relationship (\protect\includegraphics[height=1.0em]{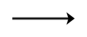}) with python scripts (\protect\includegraphics[height=1.0em]{images/script.png}). }
    \label{fig:funcs}
\end{figure}

The \textit{FA\_Estimator} script includes the main functionalities as in Fig.\ref{fig:funcs}. The \textit{monte\_carlo\_pf} and \textit{exhaustive\_pf} functions apply PF-based FA estimation algorithms. The \textit{opf} function applies the OPF-based FA estimation. The \textit{tc\_plus}, \textit{tc\_plus\_merge}, \textit{tc\_plus\_save\_tensors}, \textit{tc\_plus\_adapt} functions perform different versions of the TensorConvolution+ algorithm. The common inputs for all main functionalities are the network pandapower object (\textit{net}), the network name (\textit{net\_name}), indices of load FSPs (\textit{fsp\_load\_indices}), indices of distributed generation FSPs (\textit{fso\_dg\_indices}), scenario type for initial topology and OCs (\textit{scenario\_type}), and system constraints for maximum component loading [\%] (\textit{max\_curr\_per}), maximum voltage [p.u.] (\textit{max\_volt\_pu}), and minimum voltage [p.u.] (\textit{min\_volt\_pu}). All functionality inputs are optional. However, to estimate FAs, at least one distributed generation or load FSP is required. 
The remaining scripts, at the right of Fig.\ref{fig:funcs}, provide functions and sub-processes to implement the main functionalities.

\section{Implementation and empirical results} \label{sec:impl}

The main building blocks for the implemented FA estimation algorithms are (i) initializing network and FA estimation settings, (ii) performing simulations for FSP flexibility shifts on the network, (iii) processing the simulation results, and (iv) plotting and storing the simulation results. All functions have similar block (i), the Alg.\ref{alg:b1}. The plotting functions differ between the functionalities.

\begin{algorithm}[!tb]
\caption{Initialize network and FA estimation settings.}\label{alg:b1}
\begin{algorithmic}
\Require \textit{fsp\_load\_indices} and/or \textit{fsp\_dg\_indices},
\State initialize \textit{SettingReader} object with estimation settings,
\State check if SettingReader has acceptable values,
\If{\textit{net} is None},
    \State \textit{net} $\gets$ pandapower network with name=\textit{net\_name},
    \State change \textit{net} topology and OC for \textit{scenario\_type},
\EndIf
\State return \textit{SettingReader} with \textit{net}.
\end{algorithmic}
\end{algorithm}

\subsection{PF-based functions}

The PF-based functions differ from the FSP flexibility shift sampling functions. Therefore, the main difference between the PF-based functions is:
\begin{itemize}
    \item \textit{monte\_carlo\_pf} uses a probability \textit{distribution} (input) to obtain \textit{no\_samples} (input) of flexibility shift combinations. 
    \item \textit{exhaustive\_pf} uses the increments \textit{dp, dq} (inputs) for P and Q to sample all possible discretized flexibility shift combinations. 
\end{itemize}

Alg.\ref{alg:b345} illustrates the algorithm for both the PF-based functions after the samples are obtained. The \textit{samples} array includes the FSP shift combination samples. PCC is the point of common coupling between the TSO and the DSO.

\begin{algorithm}[!tb]
\caption{PF-based FA estimation.}\label{alg:b345}
\begin{algorithmic}
\Require \textit{samples}, \textit{SettingReader}, \textit{net},
\State \textit{init\_net} $\gets$ \textit{net},
\For{\textit{sample} $\in$ \textit{samples}},
\State apply \textit{sample} on \textit{net},
\State run PF on \textit{net},
\If{\textit{net} OC are within system constraints},
    \State store \textit{sample} index and PCC P, Q as feasible,
\Else
    \State store \textit{sample} index and PCC P, Q as not-feasible,
\EndIf
\State \textit{net} $\gets$ \textit{init\_net},
\EndFor
\State store FA PDF, CSV, and text file.
\end{algorithmic}
\end{algorithm}

\subsubsection{Empirical results}
The PF-based functions can illustrate consistent performance under various network structures and FSP combinations. However, as FSPs increase, the performance deteriorates. The Monte-Carlo-based algorithm could require large \textit{no\_samples} (e.g., $20000$) to capture the margins of the flexibility area.  
The exhaustive PF-based algorithm can become intractable for more than $3$ FSPs and small \textit{dp, dq}.

\subsection{OPF-based function}

The OPF-based algorithm applies four multi-objective optimizations (MOO). These optimizations aim to identify the maximum feasible active ($P_{PCC}$) and reactive power ($Q_{PCC}$) at the PCC achieved using the available flexibility as:
\begin{enumerate}
    \item $\mathrm{max}(\alpha P_{PCC} + (1-\alpha)Q_{PCC})$,
    \item $\mathrm{max}(-\alpha P_{PCC} + (1-\alpha)Q_{PCC})$,
    \item $\mathrm{max}(\alpha P_{PCC} + (\alpha-1)Q_{PCC})$,
    \item $\mathrm{max}(-\alpha P_{PCC} + (\alpha-1)Q_{PCC})$.
\end{enumerate}
 The variable $\alpha \in  [0, 1]$ provides a plane in which the active and reactive power shifts are combined. Therefore, the algorithm iteratively changes $\alpha$ in steps provided through the additional input the \textit{opf\_step}. For example, an \textit{opf\_step}$=0.1$ results in $11$ iterations per MOO, thus $44$ OPFs to estimate the FA.

\subsubsection{Empirical results}
The OPF-based function has convergence issues for different network structures. The OPF-based function can converge for the radial CIGRE MV network when ignoring transformer loading limitations but might not converge in other networks, e.g., the Oberrhein network.

\subsection{TensorConvolution+ functions}
The TensorConvolution+ functions correspond to the algorithm proposed in \cite{10663439}. The function \textit{tc\_plus} corresponds to the generic approach of the algorithm, whereas the rest accommodate specific use cases.

TensorConvolution+ initially creates samples of all flexibility shifts for each FSP with increments \textit{dp, dq} (inputs) for active and reactive power, respectively. The \textit{flex\_shape} input characterizes the boundaries of each FSP flexibility. Currently, the FSP shapes can be:
\begin{enumerate}
    \item \textit{Smax}: The FSP output apparent power cannot exceed its maximum apparent power, resulting in a semi-oval flexibility shape.
    \item \textit{PQmax}: The FSP active and reactive power outputs cannot exceed the maximum apparent power, resulting in a rectangular flexibility shape.
\end{enumerate}
Using these samples and the outputs of Alg.\ref{alg:b1}, the function \textit{tc\_plus} performs Alg.\ref{alg:b345tc} to estimate and plot the FA. The main differences between \textit{tc\_plus}, \textit{tc\_plus\_merge}, \textit{tc\_plus\_save\_tensors}, and \textit{tc\_plus\_adapt}:
\begin{itemize}
    \item  \textit{tc\_plus\_save\_tensors} stores extracted information and sensitivity tensors locally. This functionality reduces the tensors' memory requirements using tensor train decomposition (TTD). This reduction causes delays and is therefore excluded from the \textit{tc\_plus} function. 
    \item \textit{tc\_plus\_adapt} does not sample flexibility shifts nor estimates network component sensitives, as it adapts from the FA estimated in previous simulations for the same FSP offers.
    \item \textit{tc\_plus\_merge} is useful when memory limitations do not allow estimating FAs with the \textit{tc\_plus} function. The \textit{tc\_plus\_merge} function estimates the electrical distance between all FSPs. When a network component is sensitive to more than \textit{max\_FSPs} (input), this function merges the flexibility between the two electrically closest components iteratively until the network component is sensitive to \textit{max\_FSPs}. 
\end{itemize}

\subsubsection{Empirical results}
The TensorConvolution+ functions perform computationally better in GPUs, where tensor operations can be faster. Simulations in different network topologies showcased consistent performance with TensorConvolution+.

TensorConvolution+ can have memory issues and terminate the simulation for networks with multiple components close to the system constraints, small \textit{dp, dq}, and increased FSPs. GPUs with higher VRAM reduce these limitations. When memory issues persist, \textit{tc\_plus\_merge} can mitigate these issues but could reduce the estimation accuracy.

\begin{algorithm}[!tb]
\caption{\textit{tc\_plus} FA estimation.}\label{alg:b345tc}
\begin{algorithmic}
\Require \textit{samples}, \textit{SettingReader}, \textit{net}, 
\State \textit{init\_net} $\gets$ \textit{net},
\State $\Omega^{\gamma} \gets$ set of \textit{net} components,
\For{\textit{sample} $\in$ \textit{samples}},
    \State run PF on \textit{sample},
    \State record \textit{sample} impact on network components,
\EndFor
\State $\Omega_{sm}^{FSPs}$ $\gets$ set of FSPs with capacity smaller than $dp,dq$, 
\State \textit{impacts} $\gets$ the FSP impacts on each $\gamma \in \Omega^{\gamma}$,
\State \textit{uFA} $\gets$ the unconstrained FA using convolutions on all $FSP \in \Omega^{FSP}$,
\State $\Omega^{FSP}_{\gamma}$ $\gets$ set of FSPs that can impact each $\gamma$ more than the sensitivity thresholds,
\State $\Omega^{FSP'}_{\gamma}$ $\gets$ set of FSPs that cannot impact each $\gamma$ more than the sensitivity thresholds,
\State $\Omega^{\gamma} \gets$ remove all $\gamma$ that cannot reach the system constraints from the maximum FSP impacts from $\Omega^{\gamma}$,
\For{$\gamma \in \Omega^{\gamma}$},
    \State $\Xi_{\gamma}\gets$ apply tensor-convolution for all feasible $\Omega^{FSP}_{\gamma}$ combinations,
    \State $A_{\gamma}\gets$ sum $\Xi_{\gamma}$ in all dimensions except the first $2$,
    \State $\Upsilon_{\gamma} \gets$ apply convolution between $A_{\gamma}$ and the $\Omega^{FSP'}_{\gamma}$ shifts,
\EndFor
\If{$\Omega^{\gamma} = \emptyset$},
    \State $FA \gets uFA$,
\Else
    \State $FA \gets$ the element-wise minimum between $\Upsilon_{\gamma} \forall \gamma \in \Omega^{\gamma}$,
\EndIf
\State normalize $FA$,
\State get axes of result and create a result data frame,
\If{$\Omega_{sm}^{FSPs} \neq \emptyset$},
    \State $FA_i \gets $ bilinear interpolation on $FA$ to increase its size,
    \State $FA_{sm}\gets$ convolute $FA_i$ with the FSPs in $\Omega_{sm}^{FSPs}$,
    \State normalize $FA_{sm}$, 
\EndIf
\State plot FA and store CSV of results.
\end{algorithmic}
\end{algorithm}

\section{Illustrative examples} \label{sec:ill}
All examples were performed using the A100 GPU in Google Colab \cite{bisong2019google}. To use the package, the user can perform two steps. The first step is installing the package through pip as:
\begin{lstlisting}[language=Python]
pip install TensorConvolutionPlus
\end{lstlisting}
The second step is importing the package's \textit{FA\_Estimator} in a Python script as: 
\begin{lstlisting}[language=Python]
from TensorConvolutionPlus import FA_Estimator as TCP
\end{lstlisting}
The user can use any main function from Fig.\ref{fig:funcs} using the imported \textit{TCP}. The following subsections showcase the main functions of the package after the above steps.

\subsection{Monte Carlo PF}
\begin{figure}[!tb]
\centering
\subfigure[Uniform distribution with $6000$ samples.]{%
\label{fig:mc1}%
\begin{tikzpicture}
    \tikzstyle{every node}=[font=\scriptsize]
    \begin{axis}[
        axis on top,
        width=2.75cm,
        height=2.75cm,
        scale only axis,
        enlargelimits=false,
        xmin=17.941,
        xmax=19.035,
        ymin=5.775,
        ymax=7.225,
        axis equal=false,
        ylabel={$Q [\mathrm{MVAR}]$}, xlabel={$P [\mathrm{MW}]$},
        y label style={at={(0.15,0.5)}},
                x label style={at={(0.5,0.1)}}
        ]
      \addplot[thick,blue] graphics[xmin=17.941, xmax=19.035, ymin=5.775, ymax=7.225] {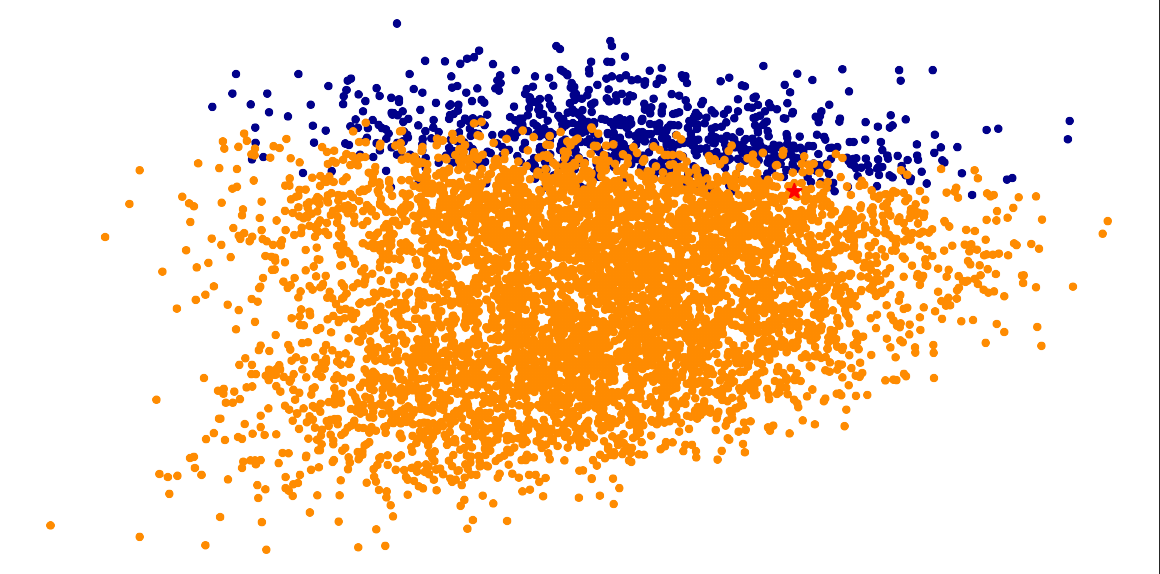};
    \end{axis}
\end{tikzpicture}}
\qquad
\subfigure[Kumaraswamy distribution with $6000$  samples.]{%
\label{fig:mc2}%
\begin{tikzpicture}
    \tikzstyle{every node}=[font=\scriptsize]
    \begin{axis}[
        axis on top,
        width=2.75cm,
        height=2.75cm,
        scale only axis,
        enlargelimits=false,
        xmin=18.023,
        xmax=18.983,
        ymin=5.385,
        ymax=7.119,
        axis equal=false,
        ylabel={$Q [\mathrm{MVAR}]$}, xlabel={$P [\mathrm{MW}]$},
        y label style={at={(0.15,0.5)}},
                x label style={at={(0.5,0.1)}}
        ]
      \addplot[thick,blue] graphics[ xmin=18.023, xmax=18.983, ymin=5.385, ymax=7.119] {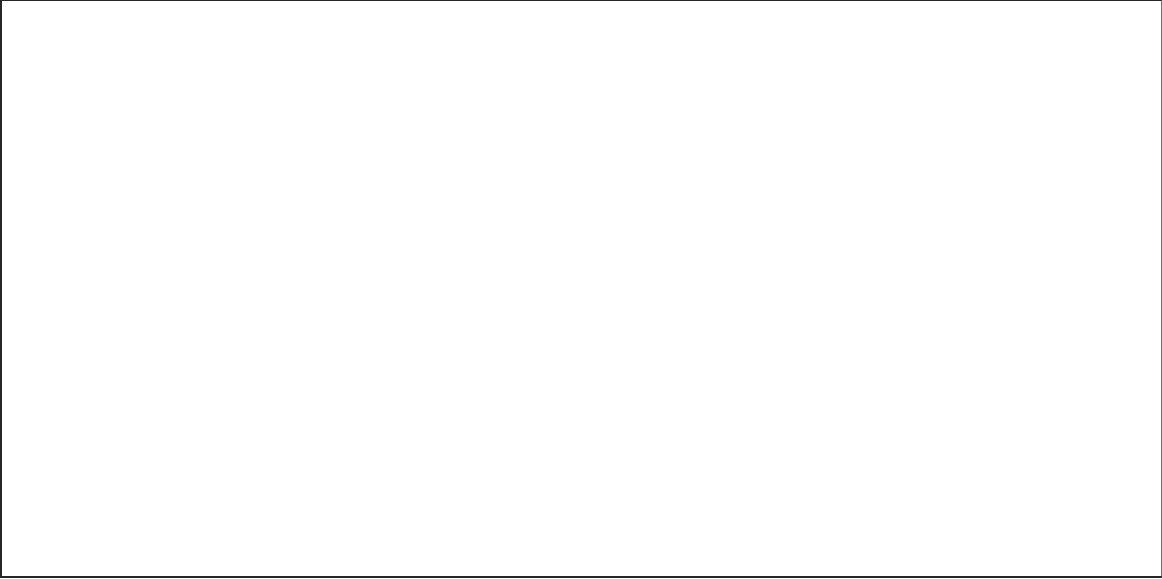};
    \end{axis}
\end{tikzpicture}%
}
\subfigure['Hard' distribution with $6000$ samples.]{%
\label{fig:mc3}%
\begin{tikzpicture}
    \tikzstyle{every node}=[font=\scriptsize]
    \begin{axis}[
        axis on top,
        width=2.75cm,
        height=2.75cm,
        scale only axis,
        enlargelimits=false,
        xmin=17.83,
        xmax=19.17,
        ymin=5.209,
        ymax=7.593,
        axis equal=false,
        ylabel={$Q [\mathrm{MVAR}]$}, xlabel={$P [\mathrm{MW}]$},
        y label style={at={(0.15,0.5)}},
                x label style={at={(0.5,0.1)}}
        ]
      \addplot[thick,blue] graphics[xmin=17.83, xmax=19.17, ymin=5.209, ymax=7.593] {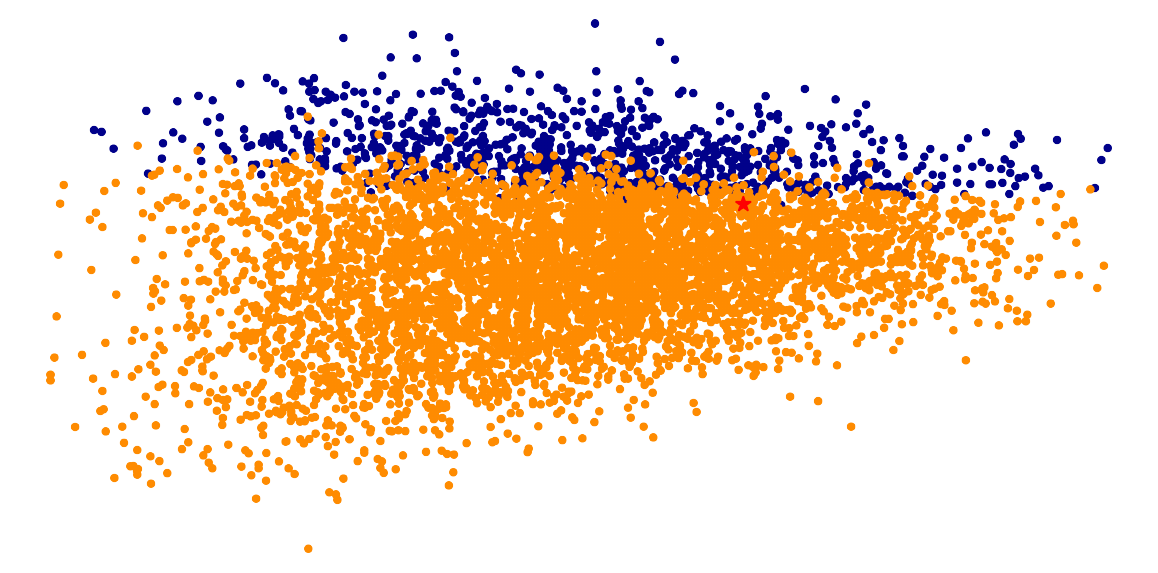};
    \end{axis}
\end{tikzpicture}
}
\qquad
\subfigure['Hard' distribution with $12000$ samples.]{%
\label{fig:mc4}%
\begin{tikzpicture}
    \tikzstyle{every node}=[font=\scriptsize]
    \begin{axis}[
        axis on top,
        width=2.75cm,
        height=2.75cm,
        scale only axis,
        enlargelimits=false,
        xmin=17.823,
        xmax=19.17,
        ymin=5.38,
        ymax=7.57,
        axis equal=false,
        ylabel={$Q [\mathrm{MVAR}]$}, xlabel={$P [\mathrm{MW}]$},
        y label style={at={(0.15,0.5)}},
                x label style={at={(0.5,0.1)}}
        ]
      \addplot[thick,blue] graphics[ xmin=17.823, xmax=19.17, ymin=5.38, ymax=7.57] {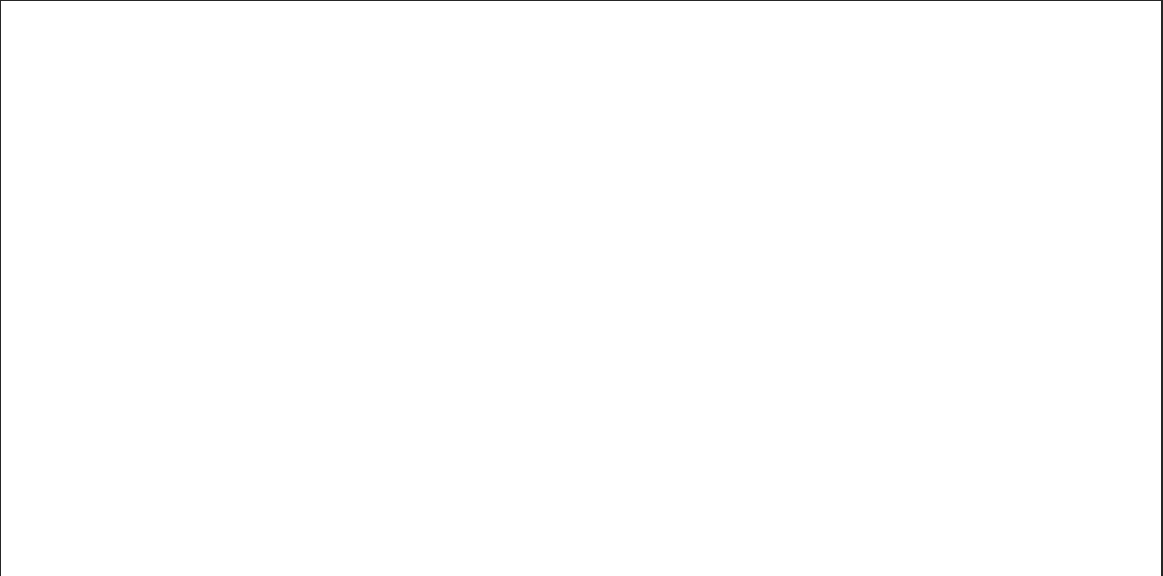};
    \end{axis}
\end{tikzpicture}%
}
\caption{Monte Carlo-based FA estimations using the 'Uniform', 'Kumaraswamy' and 'Hard' distribution of \cite{10663439}, for $6000$ or $12000$ samples.}\label{fig:mc}
\end{figure}

This section includes examples using the Monte Carlo PF estimation functionality. These examples used the Python script code:
\begin{lstlisting}[language=Python]
TCP.monte_carlo_pf(net_name='MV Oberrhein0', no_samples=6000, fsp_load_indices=[1, 2, 3], fsp_dg_indices=[1, 2, 3], distribution='Uniform')

TCP.monte_carlo_pf(net_name='MV Oberrhein0', no_samples=6000, fsp_load_indices=[1, 2, 3], fsp_dg_indices=[1, 2, 3], distribution='Kumaraswamy')

TCP.monte_carlo_pf(net_name='MV Oberrhein0', no_samples=6000, fsp_load_indices=[1, 2, 3], fsp_dg_indices=[1, 2, 3])

TCP.monte_carlo_pf(net_name='MV Oberrhein0', no_samples=12000, fsp_load_indices=[1, 2, 3], fsp_dg_indices=[1, 2, 3])
\end{lstlisting}
Each of the above lines calls the Monte Carlo PF function to estimate an FA. Fig.\ref{fig:mc} illustrates the resulting FA for each line respectively. The lines without \textit{distribution} input automatically obtain the 'Hard' distribution. In terms of computational burden the simulations required $5$ minutes and $5$ s for Fig.\ref{fig:mc1}, $5$ minutes and $6$ s for Fig.\ref{fig:mc2}, $5$ minutes and $6$ s for Fig.\ref{fig:mc3}, and $10$ minutes and $8$ s for Fig.\ref{fig:mc4} respectively. Different distributions explore the flexibility area differently. For the examples of Fig.\ref{fig:mc}, the $6000$ or $12 000$ samples do not clearly illustrate the FA margins. 

\subsection{Exhaustive PF}
\begin{figure}[!tb]
\centering
\subfigure[]{%
\label{fig:exh1}%
\begin{tikzpicture}
    \tikzstyle{every node}=[font=\scriptsize]
    \begin{axis}[
        axis on top,
        width=2.75cm,
        height=2.75cm,
        scale only axis,
        enlargelimits=false,
        xmin=17.835,
        xmax=18.901,
        ymin=5.308,
        ymax=7.546,
        axis equal=false,
        ylabel={$Q [\mathrm{MVAR}]$}, xlabel={$P [\mathrm{MW}]$},
        y label style={at={(0.15,0.5)}},
                x label style={at={(0.5,0.1)}}
        ]
      \addplot[thick,blue] graphics[xmin=17.835,
        xmax=18.901,
        ymin=5.308,
        ymax=7.546] {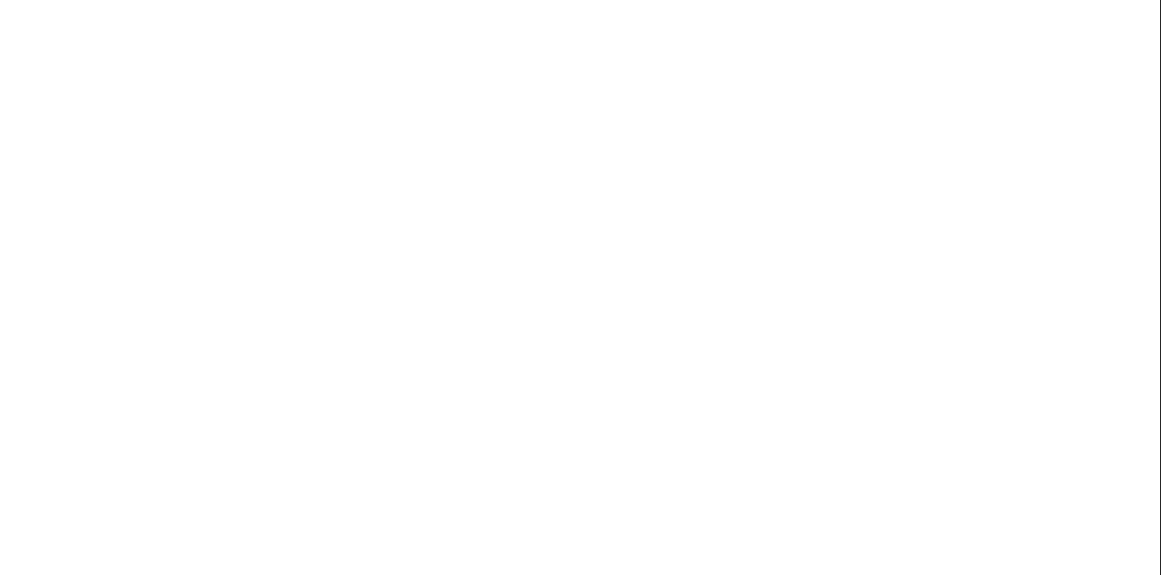};
    \end{axis}
\end{tikzpicture}}
\subfigure[]{%
\label{fig:exh2}%
\begin{tikzpicture}
    \tikzstyle{every node}=[font=\scriptsize]
    \begin{axis}[
        axis on top,
        width=2.75cm,
        height=2.75cm,
        scale only axis,
        enlargelimits=false,
        xmin=18.236,
        xmax=18.77,
        ymin=6.007,
        ymax=7.134,
        axis equal=false,
        ylabel={$Q [\mathrm{MVAR}]$}, xlabel={$P [\mathrm{MW}]$},
        y label style={at={(0.15,0.5)}},
                x label style={at={(0.5,0.1)}}
        ]
      \addplot[thick,blue] graphics[xmin=18.236,
        xmax=18.77,
        ymin=6.007,
        ymax=7.134] {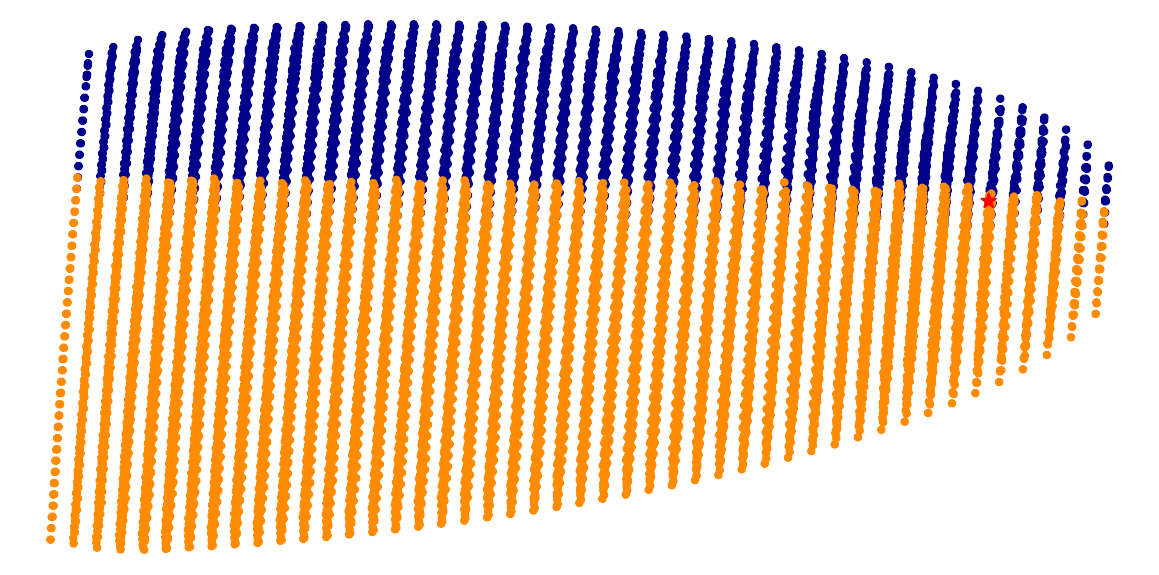};
    \end{axis}
\end{tikzpicture}}
\caption{Exhaustive Power Flow-based FA estimation with low resolution and $6$ FSPs in Fig.\ref{fig:exh1} or high resolution and $2$ FSPs in Fig.\ref{fig:exh2}.}\label{fig:exh}
\end{figure}

This section includes examples using the exhaustive power flow-based functionality. The script for the examples is:
\begin{lstlisting}[language=Python]
TCP.exhaustive_pf(net_name='MV Oberrhein0', dp=0.15, dq=0.3, fsp_load_indices=[1, 2, 3], fsp_dg_indices=[1, 2, 3])

TCP.exhaustive_pf(net_name='MV Oberrhein0', dp=0.01, dq=0.02, fsp_load_indices=[5], fsp_dg_indices=[5])
\end{lstlisting}
Fig.\ref{fig:exh} illustrates the resulting FAs from the above lines respectively. 
The two examples differ in the number of FSPs and resolutions. Fig.\ref{fig:exh1} performed $5832$ power flows and had the same network settings as in Fig.\ref{fig:mc1}, Fig.\ref{fig:mc2}, Fig.\ref{fig:mc3}, which performed $6000$ power flows. Fig.\ref{fig:exh2} performed $43121$ power flows. The GPU required $4$ minutes and $54 s$ for Fig.\ref{fig:exh1} and $36$ minutes and $18s$ for Fig.\ref{fig:exh2}.
Therefore, the Monte Carlo-based functions can be better than the exhaustive PF-based function in exploring FA margins for scenarios with more FSPs. Lowering the resolution for the exhaustive approach for producing clear FA margins can be intractable as FSPs increase.

\subsection{Optimal Power Flow}
\begin{figure}[!tb]
\centering
\subfigure[]{%
\label{fig:opf1}%
\begin{tikzpicture}
    \tikzstyle{every node}=[font=\scriptsize]
    \begin{axis}[
        axis on top,
        width=2.75cm,
        height=2.75cm,
        scale only axis,
        enlargelimits=false,
        xmin=25.966,
        xmax=46.219,
        ymin=-1.048,
        ymax=26.353,
        axis equal=false,
        ylabel={$Q [\mathrm{MVAR}]$}, xlabel={$P [\mathrm{MW}]$},
        y label style={at={(0.15,0.5)}},
                x label style={at={(0.5,0.1)}}
        ]
      \addplot[thick,blue] graphics[xmin=25.966,
        xmax=46.219,
        ymin=-1.048,
        ymax=26.353] {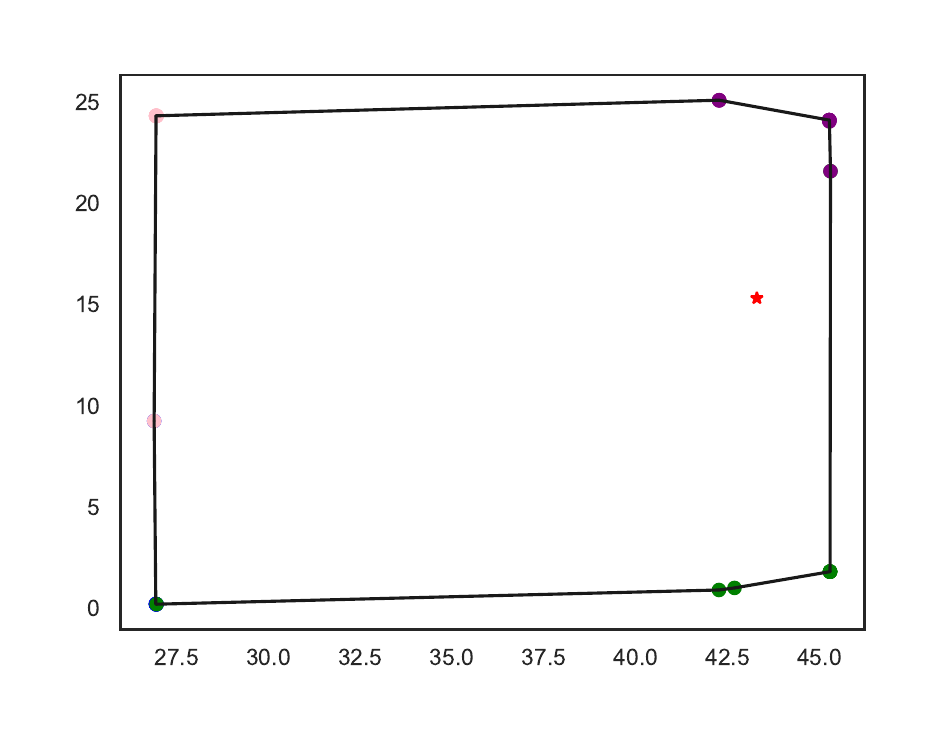};
    \end{axis}
\end{tikzpicture}}
\qquad
\subfigure[]{%
\label{fig:opf2}%
\begin{tikzpicture}
    \tikzstyle{every node}=[font=\scriptsize]
    \begin{axis}[
        axis on top,
        width=2.75cm,
        height=2.75cm,
        scale only axis,
        enlargelimits=false,
        xmin=42.066,
        xmax=45.089,
        ymin=11.832,
        ymax=16.184,
        axis equal=false,
        ylabel={$Q [\mathrm{MVAR}]$}, xlabel={$P [\mathrm{MW}]$},
        y label style={at={(0.15,0.5)}},
                x label style={at={(0.5,0.1)}}
        ]
      \addplot[thick,blue] graphics[xmin=42.066,
        xmax=45.089,
        ymin=11.832,
        ymax=16.184] {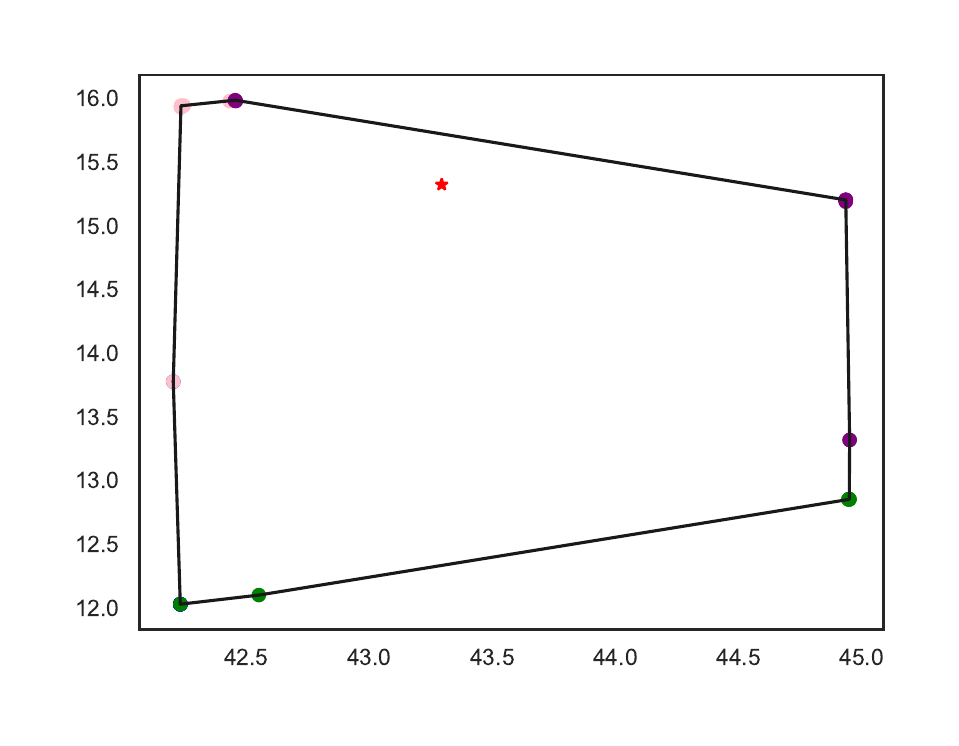};
    \end{axis}
\end{tikzpicture}}
\caption{Optimal Power Flow-based FA estimation.}\label{fig:opf}
\end{figure}

This section illustrates examples using the OPF estimation functionality. These examples used the Python script code:
\begin{lstlisting}[language=Python]
TCP.opf(net_name='CIGRE MV', opf_step=0.1, fsp_load_indices=[3, 5, 8], fsp_dg_indices=[8])

TCP.opf(net_name='CIGRE MV', opf_step=0.1, fsp_load_indices=[1, 4, 9], fsp_dg_indices=[8])
\end{lstlisting}

The script lines resulted in Fig.\ref{fig:opf} respectively. The two examples differ in the set of load FSPs. The duration for the simulations was $36.1 s$ for Fig.\ref{fig:opf1}, and $33.7s$ for Fig.\ref{fig:opf2} with $44$ converged OPFs executed in each example. Different sets of FSPs impact the FA shape as in Fig.\ref{fig:opf1}, the load FSPs had larger capacities than Fig.\ref{fig:opf2}.

\subsection{TensorConvolution+}
\begin{figure}[!tb]
\centering
\subfigure[]{%
\label{fig:tcp1}%
\begin{tikzpicture}
    \tikzstyle{every node}=[font=\scriptsize]
    \begin{axis}[
        axis on top,
        width=2.7cm,
        height=2.75cm,
        scale only axis,
        enlargelimits=false,
        xmin=17.88,
        xmax=19.28,
        ymin=5.19,
        ymax=7.99,
        axis equal=false,
        ylabel={$Q [\mathrm{MVAR}]$}, xlabel={$P [\mathrm{MW}]$},
        y label style={at={(0.2,0.5)}},
                x label style={at={(0.5,0.1)}}
        ]
      \addplot[thick,blue] graphics[xmin=17.88,
        xmax=19.28,
        ymin=5.19,
        ymax=7.99] {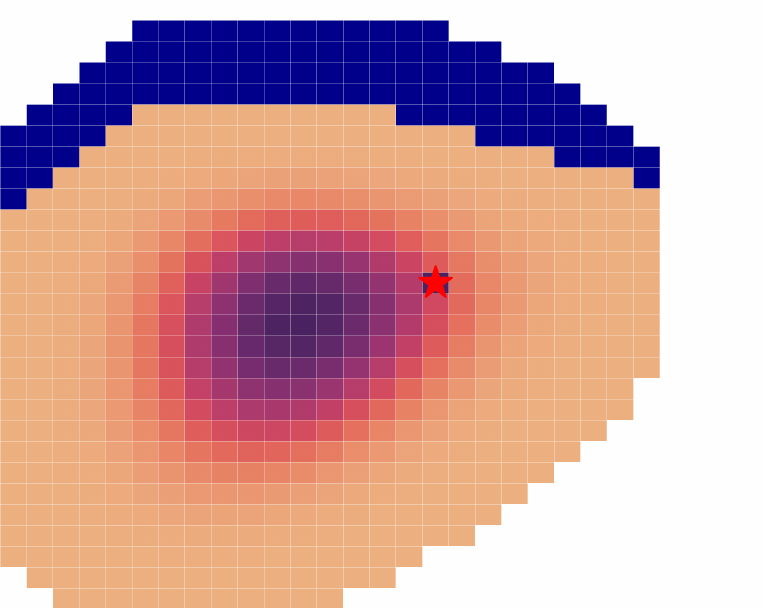};
    \end{axis}
\end{tikzpicture}%
}
\subfigure[]{%
\label{fig:tcp2}%
\begin{tikzpicture}
    \tikzstyle{every node}=[font=\scriptsize]
    \begin{axis}[
        axis on top,
        width=2.7cm,
        height=2.75cm,
        scale only axis,
        enlargelimits=false,
        xmin=18.08,
        xmax=19.23,
        ymin=5.46,
        ymax=7.96,
        axis equal=false,
        ylabel={$Q [\mathrm{MVAR}]$}, xlabel={$P [\mathrm{MW}]$},
        y label style={at={(0.2,0.5)}},
                x label style={at={(0.5,0.1)}}
        ]
      \addplot[thick,blue] graphics[xmin=18.08,
        xmax=19.23,
        ymin=5.46,
        ymax=7.96] {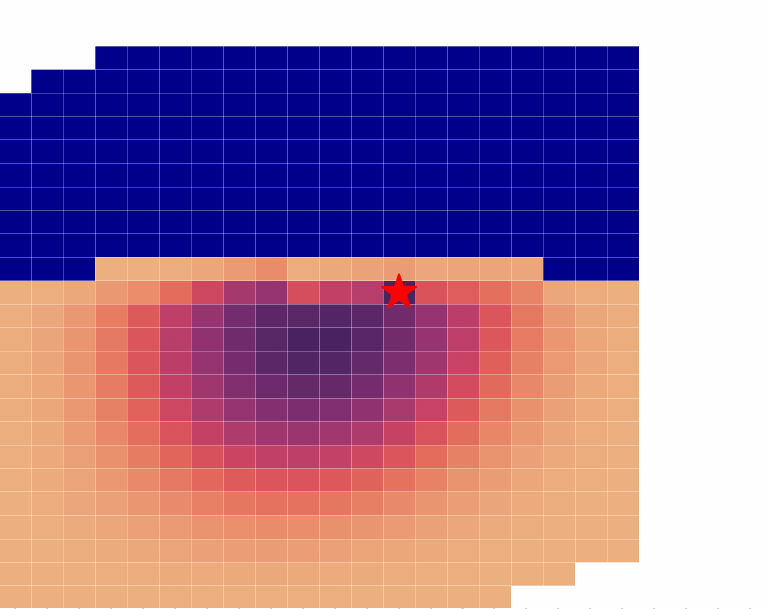};
    \end{axis}
\end{tikzpicture}
    \begin{tikzpicture}
             \tikzstyle{every node}=[font=\tiny]
    \begin{axis}[
        axis on top,
        width=1cm,
        ymode=log,
        scale only axis,
        enlargelimits=false,
        width=0.01\textwidth,
        height=0.14\textwidth,
        xmin=0,
        xmax=0.1,
        ymin=1,
        ymax=2,
        axis equal=false,
        ymajorticks=true,
        xmajorticks=false,
        xlabel={$DFC [-]$},
        ytick={1, 10^0.2, 2},
        x label style={at={(0,0.1)}, font=\scriptsize},
        ]
      \addplot[thick,blue] graphics[xmin=0, xmax=0.1, ymin=1,
        ymax=2] {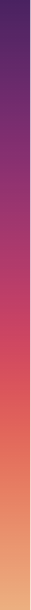};
    \end{axis}
  \end{tikzpicture}}
\caption{TensorConvolution+ algorithm examples with flex\_shape='S\_max' in Fig.\ref{fig:tcp1} and flex\_shape='PQ\_max' in Fig.\ref{fig:tcp2}.}\label{fig:tcp}
\end{figure}

This section illustrates examples using the TensorConvolution+ FA estimation functionality. The initial examples showcasing the different shapes of flexibility from FSPs used the Python lines:
\begin{lstlisting}[language=Python]
TCP.tc_plus(net_name='MV Oberrhein0', fsp_load_indices=[1, 2, 3], dp=0.05, dq=0.1, fsp_dg_indices=[1, 2, 3])

TCP.tc_plus(net_name='MV Oberrhein0', fsp_load_indices=[1, 2], dp=0.05, dq=0.1, fsp_dg_indices=[1, 2], flex_shape='PQmax')
\end{lstlisting}
The script lines resulted in Fig.\ref{fig:tcp} respectively. The example without the \textit{flex\_shape} input automatically obtains the value 'Smax'. The duration for the simulations was $13.1 s$ for Fig.\ref{fig:tcp1}, and $15s$ for Fig.\ref{fig:tcp2}. The 'Smax' shape results in flexibility areas with more oval-like shapes as in Fig.\ref{fig:tcp1}, whereas 'PQmax' results in more orthogonal shapes as in Fig.\ref{fig:tcp2}.

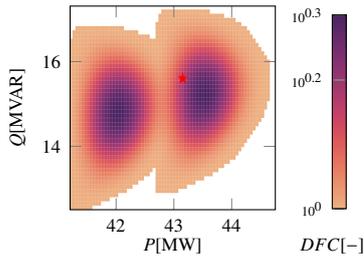
\begin{figure}[!tb]
\centering
\begin{tikzpicture}
    \tikzstyle{every node}=[font=\scriptsize]
    \begin{axis}[
        axis on top,
        width=2.7cm,
        height=2.7cm,
        scale only axis,
        enlargelimits=false,
        xmin=41.205,
        xmax=44.755,
        ymin=12.508,
        ymax=17.308,
        axis equal=false,
        ylabel={$Q [\mathrm{MVAR}]$}, xlabel={$P [\mathrm{MW}]$},
        y label style={at={(0.2,0.5)}},
                x label style={at={(0.5,0.1)}}
        ]
      \addplot[thick,blue] graphics[xmin=41.205,
        xmax=44.755,
        ymin=12.508,
        ymax=17.308] {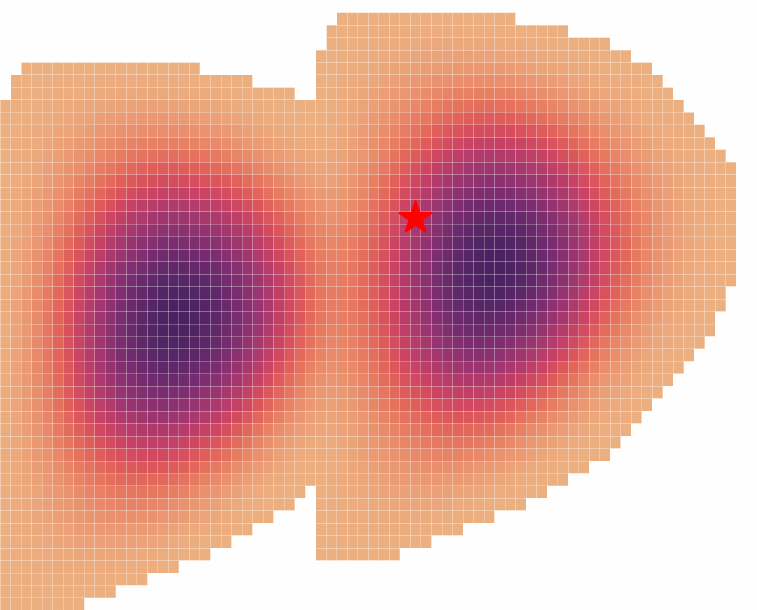};
    \end{axis}
\end{tikzpicture}
    \begin{tikzpicture}
             \tikzstyle{every node}=[font=\tiny]
    \begin{axis}[
        axis on top,
        width=1cm,
        ymode=log,
        scale only axis,
        enlargelimits=false,
        width=0.01\textwidth,
        height=0.14\textwidth,
        xmin=0,
        xmax=0.1,
        ymin=1,
        ymax=2,
        axis equal=false,
        ymajorticks=true,
        xmajorticks=false,
        xlabel={$DFC [-]$},
        ytick={1, 10^0.2, 2},
        x label style={at={(0,0.1)}, font=\scriptsize},
        ]
      \addplot[thick,blue] graphics[xmin=0, xmax=0.1, ymin=1,
        ymax=2] {images/Bars.pdf};
    \end{axis}
  \end{tikzpicture}
\caption{TensorConvolution+ algorithm example with a discrete FSP.}\label{fig:tcp3}
\end{figure}

TensorConvolution+ can also simulate FAs with FSPs, offering discrete setpoints of flexibility. For such scenarios, the input \textit{non\_linear\_fsps} specifies which of the FSPs referenced in the \textit{fsp\_dg\_indices} can only offer $2$ setpoints; current output or full output reduction. For example, a Python script could call:
\begin{lstlisting}[language=Python]
TCP.tc_plus(net_name='CIGRE MV', fsp_load_indices=[3, 4, 5], dp=0.05, dq=0.1, fsp_dg_indices=[8], non_linear_fsps=[8])
\end{lstlisting}
Fig.\ref{fig:tcp3} illustrates the result for the above function. The GPU required $17.8s$ to estimate the FA.

\subsection{TensorConvolution+ Merge}

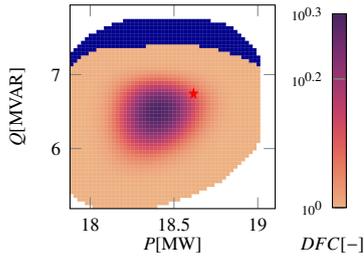
\begin{figure}[!tb]
\centering
\begin{tikzpicture}
    \tikzstyle{every node}=[font=\scriptsize]
    \begin{axis}[
        axis on top,
        width=2.7cm,
        height=2.7cm,
        scale only axis,
        enlargelimits=false,
        xmin=17.879,
        xmax=19.104,
        ymin=5.189,
        ymax=7.939,
        axis equal=false,
        ylabel={$Q [\mathrm{MVAR}]$}, xlabel={$P [\mathrm{MW}]$},
        y label style={at={(0.2,0.5)}},
                x label style={at={(0.5,0.1)}}
        ]
      \addplot[thick,blue] graphics[xmin=17.879,
        xmax=19.104,
        ymin=5.189,
        ymax=7.939] {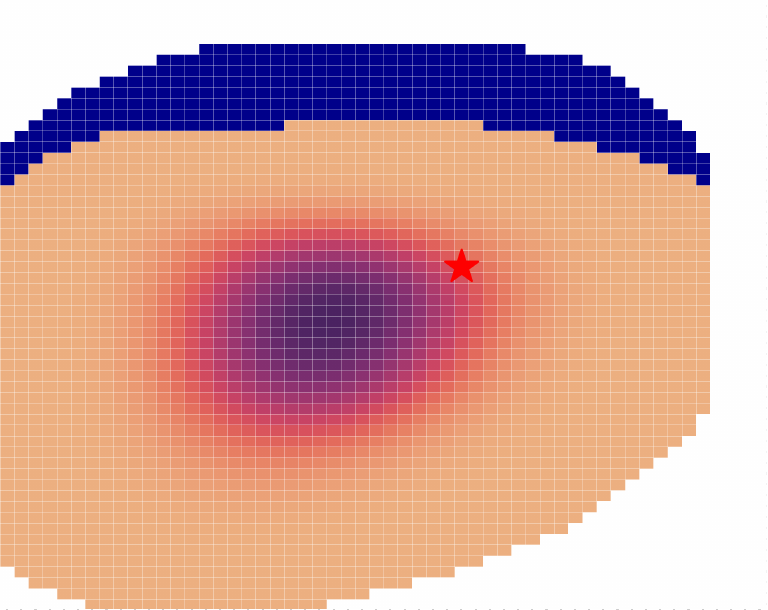};
    \end{axis}
\end{tikzpicture}
    \begin{tikzpicture}
             \tikzstyle{every node}=[font=\tiny]
    \begin{axis}[
        axis on top,
        width=1cm,
        ymode=log,
        scale only axis,
        enlargelimits=false,
        width=0.01\textwidth,
        height=0.14\textwidth,
        xmin=0,
        xmax=0.1,
        ymin=1,
        ymax=2,
        axis equal=false,
        ymajorticks=true,
        xmajorticks=false,
        xlabel={$DFC [-]$},
        ytick={1, 10^0.2, 2},
        x label style={at={(0,0.1)}, font=\scriptsize},
        ]
      \addplot[thick,blue] graphics[xmin=0, xmax=0.1, ymin=1,
        ymax=2] {images/Bars.pdf};
    \end{axis}
  \end{tikzpicture}
\caption{TensorConvolution+ algorithm example merging FSPs.}\label{fig:tcp4}
\end{figure}

This section showcases the function merging FSPs using the TensorConvolution+ algorithm. For this functionality, the \textit{max\_fsps} input determines the maximum FSPs for which a network component can be sensitive before merging their flexibility. The default value for this input is equal to the number of the load and DG FSPs minus $1$, i.e., if all FSPs impact a component's constraints, the algorithm will merge the pair with minimum electrical distance. Below is a Python script calling this function for the example of Fig.\ref{fig:tcp1} with higher resolution but maximum $5$ FSPs impactful for a component:
\begin{lstlisting}[language=Python]
TCP.tc_plus_merge(net_name='MV Oberrhein0', fsp_load_indices=[1, 2, 3], dp=0.025, dq=0.05, fsp_dg_indices=[1, 2, 3], max_fsps=5)
\end{lstlisting}
Fig.\ref{fig:tcp4} illustrates the result of the above line. The GPU needed $37.85s$.

\subsection{TensorConvolution+ Adapt}

To adapt FA estimations using estimations from expected or prior operating conditions, TensorConvolution+ requires storing the relevant information from these prior FA estimations locally. Therefore, the package's user should first call the \textit{tc\_plus\_save\_tensors} function to store the information. This function performs TTD to reduce the space required to store tensors but requires more extensive computational time than \textit{tc\_plus} to execute the additional TTD and storing operations. After \textit{tc\_plus\_save\_tensors} is executed, then \textit{tc\_plus\_adapt} can use the stored information to estimate FAs for altered related OCs if the network topology and FSPs are consistent. Bellow, an example script storing the information, altering the operating conditions, and adapting the FA for the new operating conditions:

\begin{lstlisting}[language=Python]
# Step1: Define the consistent FSPs for the storing and adapting functions
fsp_load_indices = [1, 2, 3]
fsp_dg_indices = [1, 2, 3]
# Step 2: Estimate the FA and store the relevant information for adaptation
TCP.tc_plus_save_tensors(net_name='MV Oberrhein0', fsp_load_indices=fsp_load_indices, dp=0.05, dq=0.1, fsp_dg_indices=fsp_dg_indices)
# Step 3: Modify the network operating conditions
net, net_tmp = pn.mv_oberrhein(separation_by_sub=True)
net.load['sn_mva'] = list(net.load['p_mw'].pow(2).add(net.load['q_mvar'].pow(2)).pow(0.5))
net.load['scaling'] = [1 for i in range(len(net.load))]
net.sgen['scaling'] = [1 for i in range(len(net.sgen))]
net.switch['closed'] = [True for i in range(len(net.switch))]
# Step 4: Fix the network structure 
net = fix_net(net) # This function is included in the appendix
# Step 5: Sample a new operating condition with randomness
rng = np.random.RandomState(212)
net, rng = rand_resample(net, fsp_load_indices, fsp_dg_indices, rng, 0.05, 0.01, 0.05, 0.01) # This function is also included in the appendix
# Step 6: Adapt the FA using the locally stored information
TCP.tc_plus_adapt(net=net, fsp_load_indices=fsp_load_indices, fsp_dg_indices=fsp_dg_indices)
# Step 7: Estimate the FA without adapting to compare with the above-adapted result
TCP.tc_plus(net=net, fsp_load_indices=fsp_load_indices, fsp_dg_indices=fsp_dg_indices, dp=0.05, dq=0.1)
\end{lstlisting}

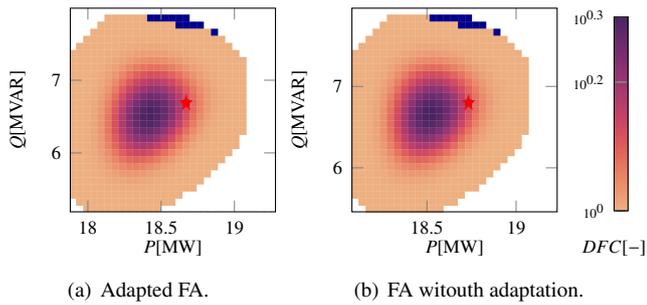
\begin{figure}[!tb]
\centering
\subfigure[Adapted FA.]{%
\label{fig:tcp5}%
\begin{tikzpicture}
    \tikzstyle{every node}=[font=\scriptsize]
    \begin{axis}[
        axis on top,
        width=2.7cm,
        height=2.7cm,
        scale only axis,
        enlargelimits=false,
        xmin=17.88,
        xmax=19.28,
        ymin=5.19,
        ymax=7.99,
        axis equal=false,
        ylabel={$Q [\mathrm{MVAR}]$}, xlabel={$P [\mathrm{MW}]$},
        y label style={at={(0.2,0.5)}},
                x label style={at={(0.5,0.1)}}
        ]
      \addplot[thick,blue] graphics[xmin=17.88,
        xmax=19.28,
        ymin=5.19,
        ymax=7.99] {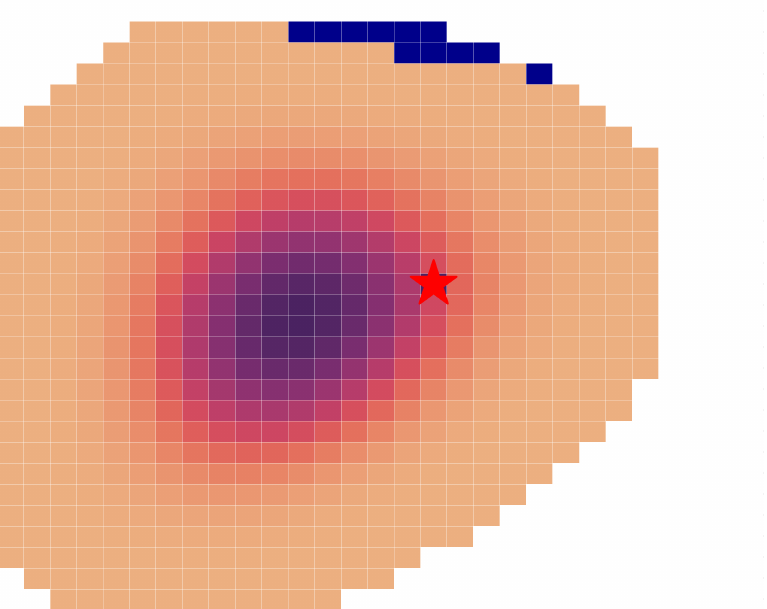};
    \end{axis}
\end{tikzpicture}%
}
\subfigure[FA witouth adaptation.]{%
\label{fig:tcp6}%
\begin{tikzpicture}
    \tikzstyle{every node}=[font=\scriptsize]
    \begin{axis}[
        axis on top,
        width=2.7cm,
        height=2.7cm,
        scale only axis,
        enlargelimits=false,
        xmin=18.08,
        xmax=19.23,
        ymin=5.46,
        ymax=7.96,
        axis equal=false,
        ylabel={$Q [\mathrm{MVAR}]$}, xlabel={$P [\mathrm{MW}]$},
        y label style={at={(0.2,0.5)}},
                x label style={at={(0.5,0.1)}}
        ]
      \addplot[thick,blue] graphics[xmin=18.08,
        xmax=19.23,
        ymin=5.46,
        ymax=7.96] {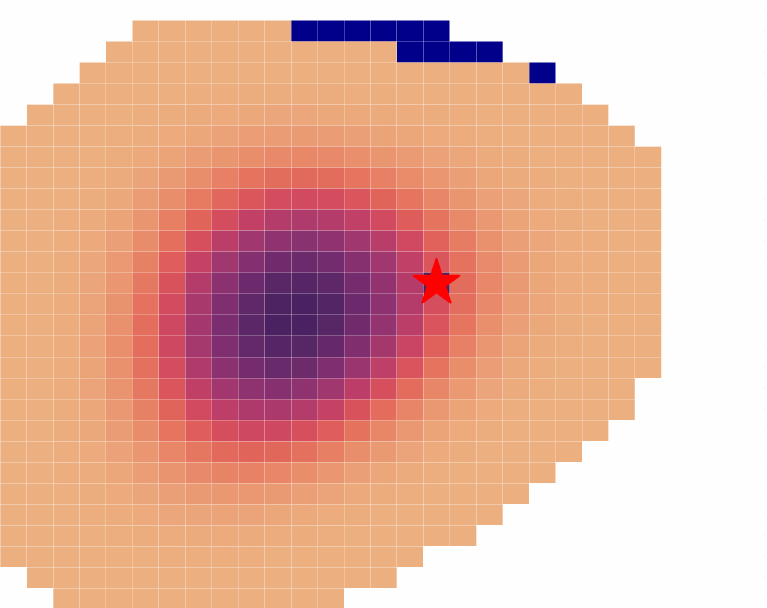};
    \end{axis}
\end{tikzpicture}
    \begin{tikzpicture}
             \tikzstyle{every node}=[font=\tiny]
    \begin{axis}[
        axis on top,
        width=1cm,
        ymode=log,
        scale only axis,
        enlargelimits=false,
        width=0.01\textwidth,
        height=0.14\textwidth,
        xmin=0,
        xmax=0.1,
        ymin=1,
        ymax=2,
        axis equal=false,
        ymajorticks=true,
        xmajorticks=false,
        xlabel={$DFC [-]$},
        ytick={1, 10^0.2, 2},
        x label style={at={(0,0.1)}, font=\scriptsize},
        ]
      \addplot[thick,blue] graphics[xmin=0, xmax=0.1, ymin=1,
        ymax=2] {images/Bars.pdf};
    \end{axis}
  \end{tikzpicture}}
\caption{TensorConvolution+ algorithm examples with flex\_shape='S\_max' in Fig.\ref{fig:tcp5} and flex\_shape='PQ\_max' in Fig.\ref{fig:tcp6}.}\label{fig:tcp7}
\end{figure}

The resulting FA from the storing function is the same as in Fig.\ref{fig:tcp1}. However, this function also stores:
\begin{enumerate}
    \item TTD results for $20$ network components with total size $241MB$.
    \item FA axes values with total size $2KB$.
    \item Matrix of the unconstrained convolution results with total size $4KB$.
    \item Dictionary with FSP impacts $382KB$.
    \item Dictionary of impactful FSPs per network component $4KB$. 
\end{enumerate}
The storing function required $61s$. Fig.\ref{fig:tcp7} illustrates the adapted FA and the FA without adaptation, i.e., not using the stored information. The FAs of Fig.\ref{fig:tcp7} have a high resemblance. The GPU needed $1.4s$ for the adapted FA of Fig.\ref{fig:tcp5} and $10.4s$ for the FA of Fig.\ref{fig:tcp6}.

\section{Impact} \label{sec:imp}
TensorConvolutionPlus is the first open-source package for FA estimation. The developed package includes different FA estimation approaches, allowing users to select and identify the best performing on their tasks. 
Nevertheless, this package focuses on the TensorConvolution+ \cite{10663439} algorithm, which can require more complex implementation compared to OPF-based and PF-based algorithms. Through this package, researchers will be able to familiarize themselves with the FA estimation topic and the TensorConvolution+ algorithm and further advance the field of FA estimation. Similarly, power system operators can use this package directly for their networks and case studies, improving the potential of adopting FAs in their operations. 

Users can execute the package functionalities with ease. The package can estimate FAs only with two lines of Python code, importing the package and calling the selected functionality. Nevertheless, the structure also allows using networks developed by the user. The software design strengthens the potential for further expansion, improvement, and adoption of FA estimation methods.

\section{Conclusions} \label{sec:conc}

Power system digitalization and developing open-source power system specialized tools are significant for intelligent and effective power grid operations. In the absence of open-source tools for FA estimation, the developed package can improve the reachability and adoption of TensorConvolution+ and FA estimation algorithms in academia and industry. With a user-friendly structure, the package allows straightforward installation and execution of FA estimation. 

The package documentation showcases example usages and details on the scripts and their functions. The package structure diversifies between FA sub-processes. This diversification allows users to better understand and expand the FA estimation algorithms in specific sub-processes, such as the FSP shapes, without impacting the remaining sub-processes. 

\section*{Acknowledgements}
This research is part of the research program 'MegaMind - Measuring, Gathering, Mining and Integrating Data for Self-management in the Edge of the Electricity System', (partly) financed by the Dutch Research Council (NWO) through the Perspectief program under number P19-25.

\bibliographystyle{elsarticle-num}

\end{document}